\RequirePackage[left]{lineno}
\documentclass[aps,prl,twocolumn,superscriptaddress,amsmath,amssymb,preprintnumbers]{revtex4}
\usepackage{amsmath}
\usepackage{mathtools}

\usepackage{graphicx} 
\usepackage{epstopdf} 
\usepackage{dcolumn} 
\usepackage{xcolor} 
\usepackage{amsmath,amssymb} 
\usepackage{hyperref} 
\usepackage[utf8]{inputenc} 
\usepackage[english]{babel} 
\usepackage{blindtext} 
\usepackage{ifthen} 
\usepackage{orcidlink} 

\graphicspath{{figures/}} 

\newboolean{articletitles}
\setboolean{articletitles}{false} 

\input{belle2-symbols}

\setboolean{articletitles}{true}

\newboolean{wordcount}
\setboolean{wordcount}{false} 

\ifthenelse{\boolean{wordcount}}%
{\usepackage{bibentry} 
 \usepackage{comment} 
 \excludecomment{align*} 
 \excludecomment{align} 
 \excludecomment{equation*} 
 \excludecomment{equation} 
 \excludecomment{eqnarray*} 
 \excludecomment{eqnarray} 
 \excludecomment{acknowledgments} 
 \def\maketitle{} 
}{}

\preprint{BELLE2-PUB-PH-2026-007}
\preprint{KEK Preprint 2026-3}
\makeatletter
\def\frontmatter@abstract@produce{%
  \par
  \preprintsty@sw{%
   \do@output@MVL{%
    \vskip\frontmatter@preabstractspace
    \vskip200\p@\@plus1fil
    \penalty-200\relax
    \vskip-200\p@\@plus-1fil
   }%
  }{%
   \addvspace{\frontmatter@preabstractspace}%
  }%
   \begingroup
    \dimen@\baselineskip
    \setbox\z@\vtop{\unvcopy\absbox}%
    \advance\dimen@-\ht\z@\advance\dimen@-\prevdepth
    \@ifdim{\dimen@>\z@}{\vskip\dimen@}{}%
   \endgroup
   \box\absbox
  \@ifx{\@empty\mini@notes}{}{\mini@notes\par}%
  \addvspace\frontmatter@postabstractspace
}%
\makeatother
\begin{document}

\pacs{
}

\title{
Search for $\Xi^0p$, $\Omega^- p$, and $\Omega^- n$ dibaryons in \Y1S and \Y2S decays at \belle
}

\ifthenelse{\boolean{wordcount}}{}{
  \author{M.~Abumusabh\,\orcidlink{0009-0004-1031-5425}} 
  \author{I.~Adachi\,\orcidlink{0000-0003-2287-0173}} 
  \author{A.~Aggarwal\,\orcidlink{0000-0002-5623-3896}} 
  \author{Y.~Ahn\,\orcidlink{0000-0001-6820-0576}} 
  \author{H.~Aihara\,\orcidlink{0000-0002-1907-5964}} 
  \author{N.~Akopov\,\orcidlink{0000-0002-4425-2096}} 
  \author{S.~Alghamdi\,\orcidlink{0000-0001-7609-112X}} 
  \author{M.~Alhakami\,\orcidlink{0000-0002-2234-8628}} 
  \author{N.~Althubiti\,\orcidlink{0000-0003-1513-0409}} 
  \author{K.~Amos\,\orcidlink{0000-0003-1757-5620}} 
% ...
\collaboration{The Belle and Belle II Collaborations}

}

\begin{abstract}
We search for $\Xi^0p$, $\Omega^-p$, and $\Omega^-n$ dibaryon states
in \Y1S and \Y2S decays, probing mass regions near the corresponding baryon-pair thresholds.
Multistrange baryon-baryon interactions are relevant to neutron-star matter but remain largely unconstrained.
Experimental and theoretical studies supporting attractive $\Xi N$ and $\Omega N$ interactions, where $N$ denotes a nucleon, 
motivate searches for weakly bound states.
We use samples of $102$ million \Y1S and $158$ million \Y2S decays collected
with the \belle detector at the KEKB asymmetric-energy $e^+e^-$ collider.
No significant signals are observed, and the first $90$\% confidence-level upper limits are set
on the branching fractions of $\Y1S$ and $\Y2S$ decays to 
$\Xi^0p$, $\Omega^-p$, and $\Omega^-n$ dibaryon states, at the level of
$\mathcal{O}(10^{-7})$--$\mathcal{O}(10^{-6})$, depending on the channel and the assumed mass difference 
from the corresponding baryon-pair threshold.
\end{abstract}

\maketitle


%
%
Bound states of two baryons with strangeness provide stringent experimental constraints on models of 
the baryon-baryon interaction, which is essential for quantifying three-body baryon forces 
relevant to modeling dense matter in neutron stars~\cite{Demorest:2010bx,Djapo:2008au,Lonardoni:2014bwa}.
Recent theoretical and experimental studies support attractive interactions in the $\Xi N$ 
and $\Omega N$ systems~\cite{ALICE:2019hdt,ALICE:2020mfd,Morita:2016auo,HALQCD:2018qyu,Sekihara:2023ihc},
where $N$ denotes a nucleon, motivating searches for weakly bound dibaryon states near threshold.

In this Letter, we report the first search for $\Xi^0p$, $\Omega^-p$, and $\Omega^-n$ dibaryon states 
in \Y1S and \Y2S decays, probing mass regions near the corresponding thresholds.
We use data collected with the Belle experiment at the KEKB asymmetric-energy \epem collider.

The baryon-baryon interaction has been studied using first-principles and 
effective-field-theory approaches, such as lattice QCD and chiral effective field theory, 
as well as phenomenological models~\cite{HALQCD:2019wsz,Haidenbauer:2018gvg,Nagels:2015slg,Sekihara:2023ihc,Fujiwara:2006yh}. 
For the $\Xi N$ system, lattice QCD and chiral effective field theory consistently 
suggest a moderately attractive interaction that is insufficient to produce a bound state, 
in agreement with measurements by ALICE~\cite{ALICE:2019hdt} of the $\Xi^-p$ correlation 
function---the two-baryon momentum correlation versus their relative momentum. 
In contrast, the extended soft-core potential, a phenomenological boson-exchange model, 
allows a stronger attractive interaction and predicts a shallow bound $\Xi N$ state
with a binding energy of order $1$ MeV~\cite{Hiyama:2019kpw}. 
The $\Omega N$ interaction is expected to be more attractive than the $\Xi N$ interaction.
Femtoscopy measurements by ALICE suggest a strong attractive interaction 
from the $\Omega^- p$ correlation function~\cite{ALICE:2020mfd},
and both lattice QCD simulations and a phenomenological approach based on the constituent quark model suggest 
an attraction sufficient to form a weakly bound $\Omega N$ state~\cite{Morita:2016auo,HALQCD:2018qyu,Sekihara:2023ihc}. 
Experimental searches sensitive to near-threshold $\Xi N$ and $\Omega N$ systems can therefore help
discriminate among theoretical descriptions and provide important input for further model development.

Decays of $\PUpsilon(nS)$ mesons produced in \epem collisions offer an environment well suited for searches 
for multistrange dibaryon production.
Owing to their gluon-rich hadronization, \Y1S and \Y2S decays have been shown to produce multibaryon final states, 
as demonstrated by the observation of antideuterons~\cite{BaBar:2014ssg,CLEO:2006zjy}.
Furthermore, Belle reported evidence for pentaquark production in \Y1S and \Y2S decays~\cite{Belle:2025pey}.
Previous searches in $\PUpsilon(nS)$ decays have placed constraints on other hypothetical multiquark and 
dibaryon states~\cite{kim:2013,BaBar:2018hpv}.
Searches for dibaryons in $\PUpsilon(nS)$ decays probe a production mechanism distinct 
from hadronic and nuclear reactions and offer sensitivity to weakly bound or unbound states 
near threshold.

%
%
We use data samples containing $102$ million \Y1S and $158$ million \Y2S decays,
corresponding to integrated luminosities of $5.75~\mathrm{fb}^{-1}$ and $24.9~\mathrm{fb}^{-1}$, respectively. 
After optimizing the selection to reconstruct the $\Xi^0p$, $\Omega^-p$, and $\Omega^-n$ dibaryon states, 
we examine invariant-mass spectra to search for structures below the baryon-pair thresholds, 
as expected for weakly bound states, 
and for near-threshold structures from unbound systems. 
Background from $e^+e^-\to q\overline{q}$ processes (with $q = u,d,s,c$) is modeled using $79.4~\mathrm{fb}^{-1}$ of data 
collected at the center-of-mass energy of $10.52~\mathrm{GeV}$,
which is $1.06$~GeV and $0.52$~GeV above the \Y1S and \Y2S, respectively. 
The analysis procedure was developed and optimized using simulated samples and 
background control regions before the signal regions were examined.
Charge-conjugate modes are implied throughout unless explicitly stated otherwise.

%
%
The Belle detector~\cite{Belle:ABASHIAN2002117} consists of a silicon vertex detector and a 50-layer central drift chamber,
an array of aerogel threshold Cherenkov counters and time-of-flight scintillation counters,
and an electromagnetic calorimeter made of CsI(Tl) crystals.
All of these subsystems are located inside a superconducting solenoid providing a $1.5$~T magnetic field,
while an iron flux-return yoke outside the coil is equipped with resistive plate chambers
to identify muons and detect $K_L^0$ mesons.

We use simulated signal and background samples to optimize selection criteria, 
to determine reconstruction and selection efficiencies, and to model distributions.
Simulated samples are generated by \evtgen and \mbox{\textsc{Pythia}}\xspace~\cite{Lange:2001uf,Sjostrand:2006za}, 
and are processed via the \belle detector simulation based on \mbox{\textsc{Geant3}}\xspace~\cite{Brun:1987ma}.
Simulated events are processed in the same way as data, using the same software~\cite{Kuhr:2018lps,b2bii}.
Signal decays are simulated assuming a phase-space model through the decays $\Y1S/\Y2S \to H^{\prime}\,\overline{B}_1\,\overline{B}_2\,4\pi\,\pi^0$,
where $H^{\prime}$ is a hypothetical dibaryon, and $\overline{B}_i\;(i=1,2)$ is an antibaryon
that balances the charge and strangeness of the dibaryon system.
A representative five-body hadronic system, $4\pi\,\pi^0$, is assumed for the additional particles 
in order to restrict the available phase space.
Representative binding energies and widths for the dibaryon signals are assumed based on theoretical expectations~\cite{Nagels:2015slg,Sekihara:2023ihc,HALQCD:2018qyu}.
For the bound-state $\Xi^0 p$ hypothesis, the dibaryon is expected to decay weakly, since there is no allowed channel
for a $\Delta S=0$ transition to a lower-mass baryon-baryon system, and its lifetime is therefore expected
to be comparable to that of the $\Xi^0$ baryon.
In contrast, the $\Omega N$ bound states are expected to decay strongly and hence to have much shorter lifetimes.
For the $\Omega N$ signals, a representative width of $5~\mathrm{MeV}/c^2$ is assumed,
following the theoretical expectation~\cite{Sekihara:2023ihc}.

%
%
Dibaryon candidates are reconstructed with a kinematic fit~\cite{Belle-IIanalysissoftwareGroup:2019dlq} 
to the full decay chain, using the topology implied by the bound or unbound-state near-threshold hypothesis. 
For the $\Xi^0 p$ analysis, the same visible final state $\pi^0 \Lambda p$ is used in both cases, 
but with different fitted topologies: a direct three-body decay $\pi^0 \Lambda p$ for the bound-state hypothesis, 
or a two-body $\Xi^0 p$ configuration with $\Xi^0 \to \pi^0 \Lambda$ for the unbound hypothesis. 
For the $\Omega N$ bound-state hypotheses, we assume a strong decay of the dibaryon system to $\Xi \Lambda$ states, 
since the invariant mass of the $\Xi \Lambda$ systems is about $0.2$~GeV$/c^2$ below the $\Omega N$ threshold. 
We therefore reconstruct the states $\Xi^0 \Lambda$ with $\Xi^0 \to \pi^0 \Lambda$ (for $\Omega^-p$) 
and $\Xi^- \Lambda$ with $\Xi^- \to \pi^- \Lambda$ (for $\Omega^-n$). 
The unbound $\Omega^- p$ hypothesis is reconstructed directly as $\Omega^- p$ with $\Omega^- \to K^- \Lambda$. 

In all cases, we reconstruct $\Lambda \to p \pi^-$; for decay chains containing a $\pi^0$, 
we reconstruct $\pi^0 \to \gamma\gamma$, and its mass is constrained in the kinematic fit. 
For all decay chains except the bound $\Xi^0 p$ hypothesis, an interaction-point constraint is applied, 
consistent with prompt production at the interaction point; this improves the vertex and mass resolution of the dibaryon candidates.

Charged-particle candidates are reconstructed from tracks using standard \belle particle-identification criteria, 
with typical identification efficiencies of about $95$\% for protons and misidentification probabilities below $4$\%.
Photon candidates from $\pi^0$ decays must have energy greater than $50\,(100)$~MeV 
in the central (forward and backward) region of the calorimeter.
To suppress combinatorial background, we require a mass window of $15$~MeV$/c^2$, 
corresponding to approximately three times the diphoton mass resolution
around the known $\pi^0$ mass~\cite{ParticleDataGroup:2024cfk}, and a momentum above $100$~MeV/$c$.
We reconstruct $\Lambda$ candidates using proton identification, vertex-quality information, 
and decay-length requirements optimized in three momentum regions~\cite{Belle:2002edu}. 
We require a mass window of $6$~MeV$/c^2$, corresponding to approximately three times the $p\pi^-$ invariant-mass resolution 
around the known $\Lambda$ mass~\cite{ParticleDataGroup:2024cfk}.
The invariant-mass distributions of $\Lambda \pi^0$, $\Lambda \pi^-$ and $\Lambda K^-$ combinations 
are fitted to determine the experimental mass resolutions.
We then apply selection windows of $6$~MeV$/c^2$ for $\Xi^{0}$
and $4$~MeV$/c^2$ for $\Xi^-$ and $\Omega^-$, corresponding to approximately two standard deviations 
around the known masses~\cite{ParticleDataGroup:2024cfk}.
Since two proton tracks are required in the final state, we check for duplicate-track assignments. 
In both data and simulation, a non-negligible contribution is found from duplicate proton tracks with a back-to-back topology. 
To suppress this background, events with two proton candidates having a momentum difference smaller 
than $0.1~\mathrm{GeV}/c$ and $C^{\prime}<-5$ are rejected, 
where $C^{\prime}=\log[(\cos\theta+1)/(1-\cos\theta)]$ and $\theta$ is the opening angle between their momenta.
Selection requirements on the kinematic-fit quality and on topological variables, 
such as hyperon flight distances, are optimized using the Punzi figure of merit~\cite{Punzi:2003bu} 
for a target signal significance of $3\sigma$.
The overall reconstruction and selection efficiencies range from about $0.5$\% to $5$\%, depending on the decay channel, 
with the lowest efficiencies observed for channels involving a $\pi^0$.
The efficiencies are found to be nearly constant as a function of the assumed mass for the bound states, 
and show a small increase with the assumed mass for the unbound states~\cite{supplemental}.
About $10$\% of the events contain multiple candidates, 
with an average candidate multiplicity of less than $1.15$, 
and all candidates are retained in the analysis.
\begin{figure}[!t]
  \begin{center}
    \includegraphics[width=0.5\textwidth]{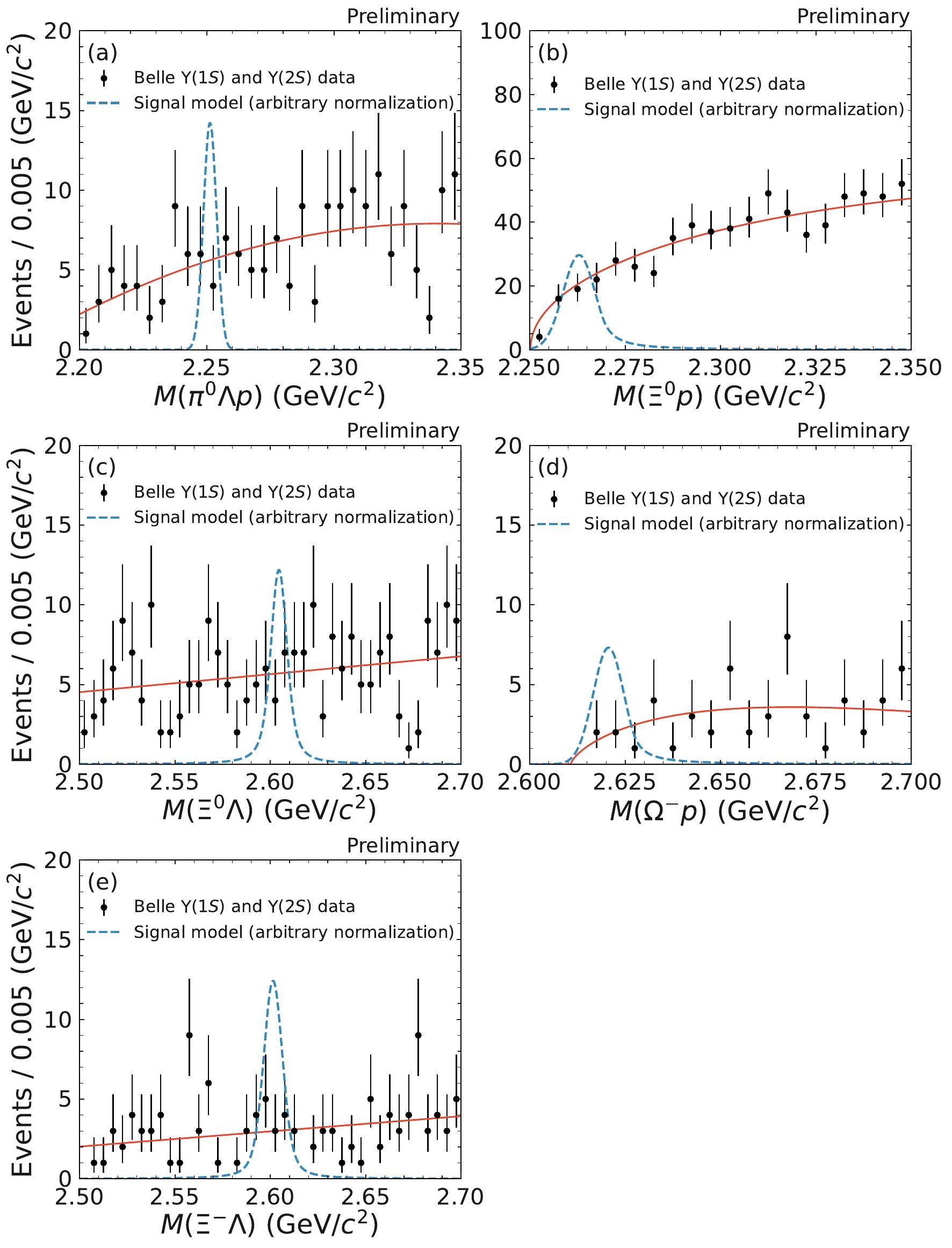}
    \vspace{-1.0em}
    \caption{Invariant-mass spectra for (left column) bound-state 
    and (right column) unbound-state hypotheses: \\
    (a) $(\Xi^0 p) \to \pi^0 \Lambda p$, \:
    (b) $(\Xi^0 p) \to \Xi^0 p$, \:
    (c) $(\Omega^- p) \to \Xi^0 \Lambda$,\\
    (d) $(\Omega^- p) \to \Omega^- p$, and
    (e) $(\Omega^- n) \to \Xi^- \Lambda$.
    Data include both \Y1S and \Y2S samples.
    Fits with the background component are overlaid as red curves.
    Blue dashed curves indicate the expected signal shapes (with arbitrary normalization),
    assuming a binding energy of $2$~MeV for the bound $\Xi^{0}p$ hypothesis, 
    $10$~MeV for the bound $\Omega^{-}p$ and $\Omega^{-}n$ hypotheses, and
    $10$~MeV above the mass threshold for the unbound hypotheses.}
    \vspace{-2.0em}
    \label{fig:y12s2xip_omp_omn_pap_bub_invm_data_w_fit_bg_only_w_signal_model}
  \end{center}
\end{figure}

%
%
Figure~\ref{fig:y12s2xip_omp_omn_pap_bub_invm_data_w_fit_bg_only_w_signal_model} shows invariant-mass spectra for the bound-state 
and the unbound-state candidates.
The \Y1S and \Y2S samples are combined, since both states decay predominantly via the same three-gluon intermediate state, 
leading to similar hadronization dynamics, and because the substantial feed-down contribution from $\Y2S \to \pi\pi \Y1S$ prevents 
a separation of inclusive \Y2S decays with the present dataset.

The invariant-mass spectra are modeled as the sum of a signal component and a combinatorial background.
We model the signal component with a Gaussian function for the bound $\Xi^{0}p$ hypothesis, 
a Breit--Wigner function convolved with a Gaussian function for the bound $\Omega^{-}p$ and $\Omega^{-}n$ hypotheses,
and a double-sided Crystal Ball function~\cite{gaiser_skwarnicki} for the unbound hypotheses.
The corresponding mass-resolution parameter varies within a few MeV$/c^2$ over the scanned mass range,
and is determined separately for the bound-state and near-threshold hypotheses
to account for their different kinematic behavior.
The model also includes a component for misreconstructed signal decays, 
whose relative contribution is determined from simulated samples. 
This contribution is sizable for channels containing a $\pi^0$, reaching about $20$\%, 
while it is below $2$\% for channels with fully charged final states. 
The misreconstructed components are modeled with analytic functions determined from simulation,
including asymmetric generalized Gaussian and Novosibirsk functions~\cite{Belle:1999bhb}, as well as Argus-like threshold shapes, 
depending on the channel and hypothesis.
The background is modeled with a first- or second-order polynomial for the bound-state hypotheses and 
with an Argus-like threshold function for the unbound near-threshold hypotheses.
All shape parameters are fixed from simulation-based studies, 
while the signal and background yields are allowed to float in the fits.

%
%
Unbinned extended maximum likelihood fits are performed to the invariant mass spectra of all five final states. 
No statistically significant structures consistent with dibaryon production are observed in any channel.
We therefore set $90$\% confidence-level upper limits on the branching fractions of \Y1S and \Y2S decays 
to these dibaryon systems, as a function of the mass difference from the corresponding baryon-pair threshold, 
using the $q_{\mu}$ test statistic~\cite{Cowan:2010js} based on the profile likelihood,
implemented in the \texttt{hepstats} package of the Scikit-HEP project~\cite{Rodrigues:2020syo}.
The branching fraction for a signal decay is calculated as
\begin{equation}
  \mathcal{B} = \frac{N_{\mathrm{sig}}}{2\,N_{\PUpsilon}\,\varepsilon_{\mathrm{rec}}\, \mathcal{B}_{\mathrm{sub}}},
\end{equation}
where $N_{\mathrm{sig}}$ is the signal yield obtained from the fit, $N_{\PUpsilon}$
is the total number of $\Y1S$ and $\Y2S$ decays, $\varepsilon_{\mathrm{rec}}$
is the reconstruction efficiency, and $\mathcal{B}_{\mathrm{sub}}$
is the product of the branching fractions of intermediate hyperon decays,
such as $\Lambda \to p\pi^-$ and $\Omega^- \to \Lambda K^-$.
The mass difference is scanned from $-30$~MeV to $30$~MeV in steps of $2$~MeV,
covering a range significantly larger than typical theoretical expectations and 
allowing direct comparison with previous searches~\cite{kim:2013}.
This scan corresponds to a mass range of about $2.22$--$2.25$~GeV/$c^2$ for the bound $\Xi^0 p$ channel,
$2.58$--$2.61$~GeV/$c^2$ for the bound $\Omega^- p$ and $\Omega^- n$ channels,
$2.25$--$2.28$~GeV/$c^2$ for the unbound $\Xi^0 p$ channel,
and $2.61$--$2.64$~GeV/$c^2$ for the unbound $\Omega^- p$ channel.

%
%
Systematic uncertainties affecting the upper limits are summarized in Table~\ref{tab:systematic_uncertainties}. 
They are incorporated as Gaussian-constrained nuisance parameters in the profile-likelihood 
construction and account for uncertainties in the fitted signal yield, 
reconstruction efficiency, number of \Y1S and \Y2S decays, and intermediate branching fractions.
The systematic uncertainty on the $\Lambda$ selection efficiency is evaluated based on a previous \belle study 
using $B \to \Lambda \overline{\Lambda} K$ decays~\cite{Belle:2016tai}.
The uncertainty associated with the fit model is estimated by varying the order of the background polynomial, 
the fit mass range in the bound-state analyses, 
and the shape parameters of the fit model within their uncertainties.
For the unbound near-threshold channels, only the fit mass range and the shape parameters are varied, 
since no alternative polynomial orders are applicable. 
The resulting variations in the signal yield are treated as systematic uncertainties.
The systematic uncertainty on the $\pi^0$ selection efficiency is evaluated based on a previous \belle study 
using $\tau^- \to \pi^- \pi^0 \nu_{\tau}$ decays~\cite{Belle:2019jeu}.
A dedicated analysis measures the number of \Y1S and \Y2S decays in the sample: 
the uncertainty on this number is included as a systematic~\cite{Belle:2016yfw}. 
We propagate uncertainties on the branching fractions of $\Lambda$ and $\Omega$ decays~\cite{ParticleDataGroup:2024cfk}. 
The assumed signal model impacts the reconstruction efficiency.
We assign a corresponding systematic uncertainty from the variation in efficiency obtained 
when considering all antibaryon combinations allowed by charge and strangeness conservation.
Systematic uncertainties associated with particle-identification and tracking efficiencies
are evaluated using a control sample of $D^{\ast +} \to D^0 \pi^+$ decays, with $D^0 \to K^- \pi^+$~\cite{Belle:2016tai}.
The resulting overall systematic uncertainties are 
at the level of $5$--$6$\% for all bound and unbound dibaryon channels, 
obtained by summing individual contributions in quadrature.

\begin{figure}[!t]
  \begin{center}
    \includegraphics[width=0.48\textwidth]{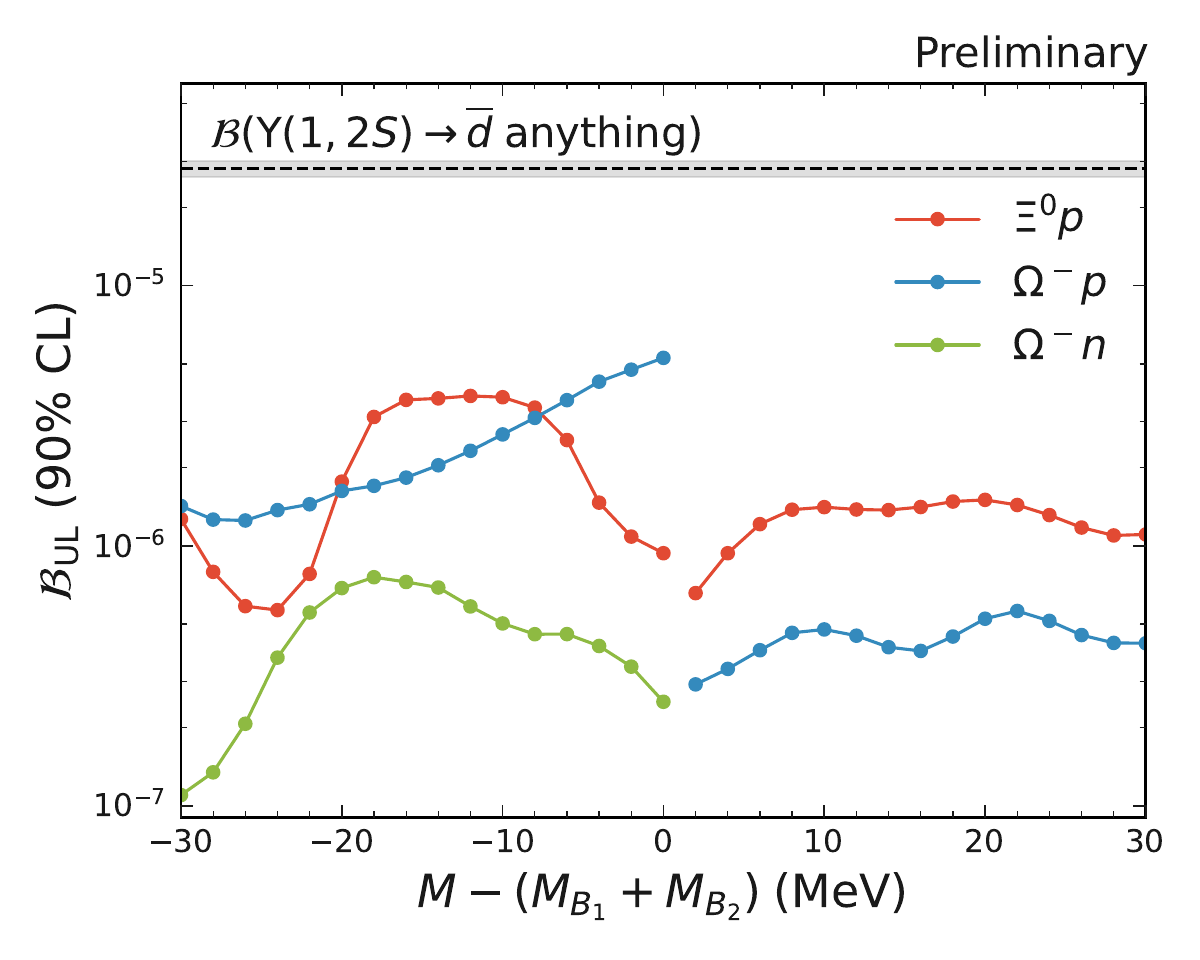}
    \vspace{-1.0em}
    \caption{Upper limits at $90$\% confidence level
    as a function of the mass difference from the corresponding baryon-pair threshold,
    where $M_{B_1}$ and $M_{B_2}$ are the masses of the first and second baryons, respectively.
    The red, blue, and green markers correspond to $\Xi^0 p$, $\Omega^- p$, and $\Omega^- n$, respectively.
    An average value of the branching fraction for antideuteron~($\overline{d}$) production 
    in \Y1S and \Y2S decays
    is indicated by the horizontal dashed line with the shaded band~\cite{ParticleDataGroup:2024cfk}.}
    \vspace{-2.0em}
    \label{fig:mass_dependent_ul_5x4}
  \end{center}
\end{figure}
Figure~\ref{fig:mass_dependent_ul_5x4} shows the upper limits as a function of
the mass difference from the corresponding baryon-pair threshold.
Assuming $C\!P$ symmetry, baryon and antibaryon channels are combined to improve the sensitivity.
For reference, the measured branching fraction for antideuteron production in \Y1S and \Y2S decays
is shown as a benchmark for the achieved sensitivity, which is comparable to that of previous dibaryon searches.
The observed upper limits are found to be consistent with the expected limits 
within statistical fluctuations for all channels~\cite{supplemental}.
For the $\Omega^- n$ system, upper limits are derived only for the bound-state hypothesis, 
since the unbound near-threshold configuration cannot be reconstructed without direct neutron detection.
The gap at zero mass difference in the $\Omega^- p$ channel reflects the reduced
reconstruction efficiency of the bound-state hypothesis, which includes a $\pi^0$
in the final state, relative to the all-charged unbound final state.
Best-fit estimates of the branching fractions for all channels are provided in the Supplemental Material~\cite{supplemental}.

%
%

In summary, we report the first search for $\Xi^0p$, $\Omega^-p$, and $\Omega^-n$ dibaryons
in \Y1S and \Y2S decays. 
We analyze $102$ million \Y1S and $158$ million \Y2S decays collected with the Belle experiment 
and find no significant signal in any channel.  
Upper limits on the corresponding branching fractions are set at the 90\% confidence level,
in the range of $\mathcal{O}(10^{-7})$ to $\mathcal{O}(10^{-6})$, depending on the channel.
These results provide new experimental constraints on the formation of multistrange dibaryons in gluon-rich bottomonium decays and 
complement existing searches in hadronic, nuclear, and heavy-ion environments.

\begin{table}[!t]
  \caption{Fractional systematic uncertainties (in percent).}
  \label{tab:systematic_uncertainties}
  \begin{center}

    \begin{tabular*}{\linewidth}{@{\extracolsep{\fill}}lccccc}
    \hline \hline
            & \multicolumn{3}{c}{Bound} & \multicolumn{2}{c}{Unbound} \\
    Source  & $\Xi^0p$ & $\Omega^-p$ & $\Omega^-n$ & $\Xi^0p$ & $\Omega^-p$ \\
    \hline
    $\Lambda$ selection efficiency         & $3.2$   & $4.1$  & $5.1$  & $3.6$  & $4.0$   \\ 
    Fit model                        & $3.1$   & $3.4$  & $0.6$  & $2.4$  & $1.1$   \\ 
    $\pi^0$ selection efficiency               & $2.2$   & $2.2$  & $-$    & $2.2$  & $-$     \\  
    Number of \Y1S and \Y2S                  & $1.5$   & $1.5$  & $1.5$  & $1.5$  & $1.5$   \\ 
    ${\cal{B}}(\Lambda \to p \pi^-)$ & $0.8$   & $1.1$  & $1.1$  & $0.8$  & $0.8$   \\ 
    ${\cal{B}}(\Omega^{-} \to \Lambda K^{-})$ & $-$   & $-$  & $-$  & $-$  & $1.0$   \\ 
    Reconstruction efficiency   & $0.6$   & $0.7$  & $0.2$  & $0.3$  & $0.1$   \\ 
    Tracking efficiency                 & $0.35$  & $-$    & $0.35$ & $0.35$ & $0.7$   \\
    Particle identification          & $0.26$  & $-$    & $1.4$  & $0.25$ & $0.38$  \\
    \hline
    Total                            & $5.3$ & $6.1$ & $5.6$ & $5.2$ & $4.6$        \\
    \hline \hline    
    \end{tabular*}
  \end{center}
  \vspace{-2.0em}
\end{table}

\begin{acknowledgments}
This work, based on data collected using the Belle detector, which was
operated until June 2010, was supported by 
the Ministry of Education, Culture, Sports, Science, and
Technology (MEXT) of Japan, the Japan Society for the 
Promotion of Science (JSPS), and the Tau-Lepton Physics 
Research Center of Nagoya University; 
the Australian Research Council including grants
DP210101900, 
DP210102831, 
DE220100462, 
LE210100098, 
LE230100085; 
Austrian Federal Ministry of Education, Science and Research (FWF) and
FWF Austrian Science Fund No.~P~31361-N36;
National Key R\&D Program of China under Contract No.~2022YFA1601903,
National Natural Science Foundation of China and research grants
No.~11575017,
No.~11761141009, 
No.~11705209, 
No.~11975076, 
No.~12135005, 
No.~12150004, 
No.~12161141008, 
and
No.~12175041, 
and Shandong Provincial Natural Science Foundation Project ZR2022JQ02;
the Czech Science Foundation Grant No. 22-18469S;
Horizon 2020 ERC Advanced Grant No.~884719 and ERC Starting Grant No.~947006 ``InterLeptons'' (European Union);
the Carl Zeiss Foundation, the Deutsche Forschungsgemeinschaft, the
Excellence Cluster Universe, and the VolkswagenStiftung;
the Department of Atomic Energy (Project Identification No. RTI 4002), the Department of Science and Technology of India,
and the UPES (India) SEED finding programs Nos. UPES/R\&D-SEED-INFRA/17052023/01 and UPES/R\&D-SOE/20062022/06; 
the Istituto Nazionale di Fisica Nucleare of Italy; 
National Research Foundation (NRF) of Korea Grants
No.~2021R1-F1A-1064008,
No.~2022R1-A2C-1003993,
No.~RS-2018-NR031074,
No.~RS-2021-NR060129,
No.~RS-2024-00354342
No.~RS-2025-02219521,
No.~RS-2026-25471491,
No.~RS-2026-25480677,
and
No.~RS-2026-25486791,
Radiation Science Research Institute,
Foreign Large-Size Research Facility Application Supporting project,
the Global Science Experimental Data Hub Center, the Korea Institute of Science and
Technology Information (K26L1M2C3) and KREONET/GLORIAD;
the Polish Ministry of Science and Higher Education and 
the National Science Center;
the Ministry of Science and Higher Education of the Russian Federation
and the HSE University Basic Research Program, Moscow; 
University of Tabuk research grants
S-1440-0321, S-0256-1438, and S-0280-1439 (Saudi Arabia);
the Slovenian Research Agency Grant Nos. J1-50010 and P1-0135;
Ikerbasque, Basque Foundation for Science, and the State Agency for Research
of the Spanish Ministry of Science and Innovation through Grant No. PID2022-136510NB-C33 (Spain);
the Swiss National Science Foundation; 
the Ministry of Education and the National Science and Technology Council of Taiwan;
and the United States Department of Energy and the National Science Foundation.
These acknowledgements are not to be interpreted as an endorsement of any
statement made by any of our institutes, funding agencies, governments, or
their representatives.
We thank the KEKB group for the excellent operation of the
accelerator; the KEK cryogenics group for the efficient
operation of the solenoid; and the KEK computer group and the Pacific Northwest National
Laboratory (PNNL) Environmental Molecular Sciences Laboratory (EMSL)
computing group for strong computing support; and the National
Institute of Informatics, and Science Information NETwork 6 (SINET6) for
valuable network support.

\end{acknowledgments}

\section*{Data Availability}
The data underlying this analysis are available from the \belle and \belletwo Collaborations upon reasonable request.

\ifthenelse{\boolean{wordcount}}%
{ \nobibliography{references} }
{ \bibliographystyle{belle2}
  \bibliography{references} }

  \cleardoublepage
\appendix
\setcounter{figure}{0}
\setcounter{table}{0}
\setcounter{page}{1}

\onecolumngrid
\renewcommand{\thefigure}{\arabic{figure}}
\renewcommand{\thetable}{\arabic{table}}
\renewcommand{\thepage}{\arabic{page}}

\subsection*{Supplemental material}

\begin{figure}[!hb]
  \begin{center}
   \includegraphics[width=0.5\textwidth]{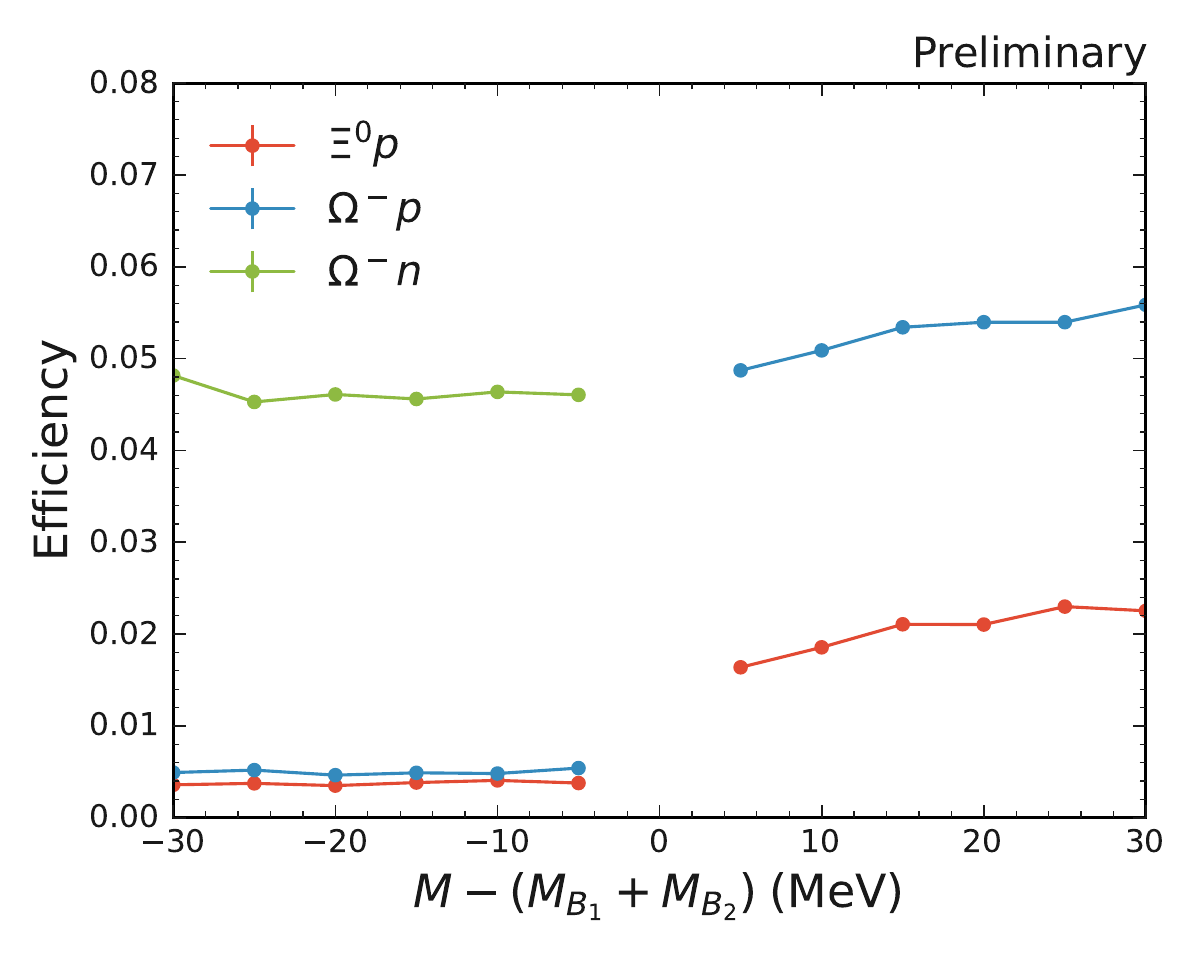}
    \caption{Reconstruction and selection efficiency for all channels as a function of 
    the mass difference from the corresponding baryon-pair threshold,
    where $M_{B_1}$ and $M_{B_2}$ are the masses of the first and second baryons, respectively.
      Statistical uncertainties are included, but they are smaller than $0.1$\% and thus not visible at the scale of the figure.}
    \label{fig:mass_dependent_eff_5x4}
  \end{center}
\end{figure}

Figure~\ref{fig:mass_dependent_eff_5x4} shows the reconstruction and selection efficiency for all channels
as a function of the mass difference from the corresponding baryon-pair threshold.

\clearpage

\begin{figure}[!hb]
  \begin{center}
    \includegraphics[width=0.7\textwidth]{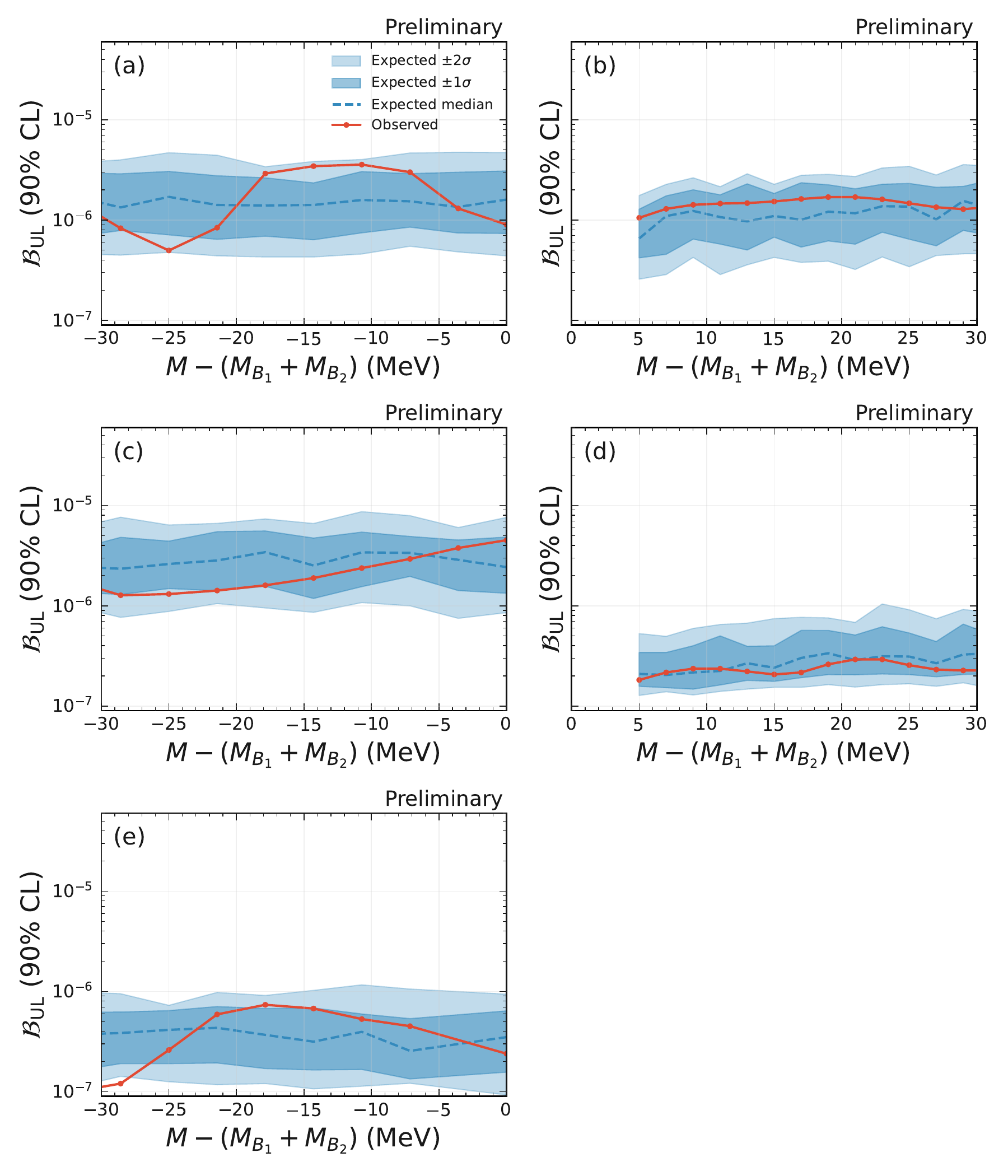}
    \caption{Observed upper limits on the branching fraction for all channels and 
    the expected limits with the $1\sigma$ and $2\sigma$ uncertainty bands
    for (left column) bound-state 
    and (right column) unbound-state hypotheses: \\
    (a) $(\Xi^0 p) \to \pi^0 \Lambda p$, 
    (b) $(\Xi^0 p) \to \Xi^0 p$, 
    (c) $(\Omega^- p) \to \Xi^0 \Lambda$,
    (d) $(\Omega^- p) \to \Omega^- p$, and 
    (e) $(\Omega^- n) \to \Xi^- \Lambda$.}
    \label{fig:brazil_band_plot}
  \end{center}
\end{figure}

Figure~\ref{fig:brazil_band_plot} shows the observed upper limits on the branching fraction for all channels 
and the expected limits with the $1\sigma$ and $2\sigma$ uncertainty bands.
All observed upper limits are consistent with those expected within $2\sigma$.

\clearpage

\begin{figure}[!hb]
  \begin{center}
    \includegraphics[width=0.5\textwidth]{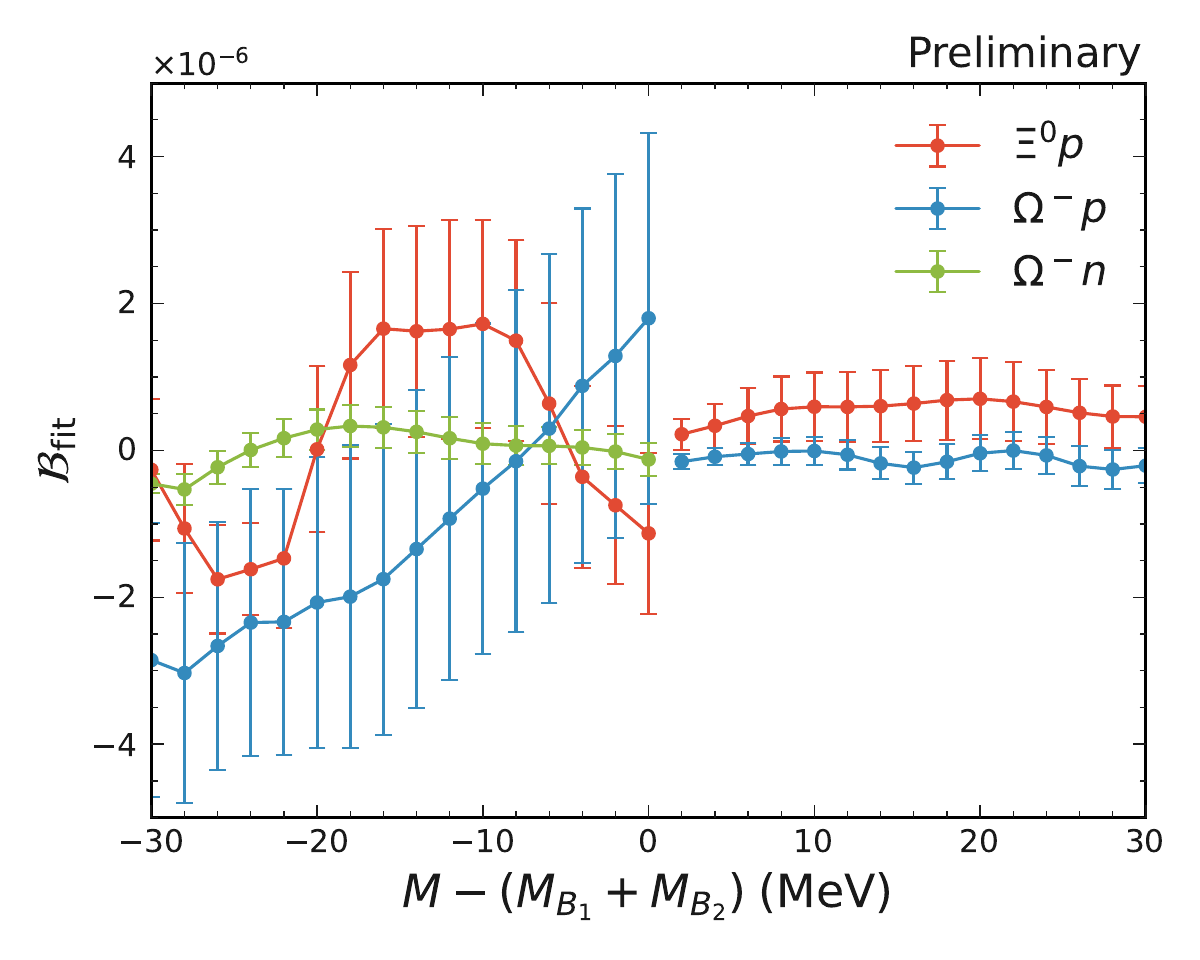}
    \caption{Best-fit branching fractions of $\Y1S$ and $\Y2S$ decays to 
    $\Xi^0p$, $\Omega^-p$, and $\Omega^-n$ dibaryon states 
    as a function of the mass difference from the corresponding baryon-pair threshold for all analysis channels.}
    \label{fig:mass_dependent_br_best_5x4}
  \end{center}
\end{figure}

Figure~\ref{fig:mass_dependent_br_best_5x4} shows the best-fit values of the
branching fractions of $\Y1S$ and $\Y2S$ decays to 
$\Xi^0p$, $\Omega^-p$, and $\Omega^-n$ dibaryon states as a
function of the mass difference from the corresponding baryon-pair threshold for all channels.
Tables~\ref{tab:best_fit_br_bound} and \ref{tab:best_fit_br_unbound} list the best fit values 
of the branching fractions for bound and unbound state hypotheses, respectively.
Channels containing a $\pi^0$ exhibit larger uncertainties because of their lower
reconstruction efficiencies than those reconstructed entirely from charged particles.

\clearpage

\begin{table}[!t]
  \caption{Best-fit values of the branching fractions for bound-state hypotheses.
  Uncertainties are statistical only.}
  \label{tab:best_fit_br_bound}
  \begin{center}
\begin{tabular}{rrrr}
  \hline \hline
 Mass difference (MeV) & $\Xi^0p$ ($\times 10^{-6}$) & $\Omega^-p$ ($\times 10^{-6}$) & $\Omega^-n$ ($\times 10^{-6}$) \\
 \hline
$-30$ & $-0.3 \pm 1.0$ & $-2.9 \pm 1.9$ & $-0.5 \pm 0.1$ \\
$-28$ & $-1.1 \pm 0.9$ & $-3.0 \pm 1.8$ & $-0.5 \pm 0.2$ \\
$-26$ & $-1.8 \pm 0.7$ & $-2.7 \pm 1.7$ & $-0.2 \pm 0.2$ \\
$-24$ & $-1.6 \pm 0.6$ & $-2.3 \pm 1.8$ & $0.0 \pm 0.2$ \\
$-22$ & $-1.5 \pm 0.9$ & $-2.3 \pm 1.8$ & $0.2 \pm 0.3$ \\
$-20$ & $0.0 \pm 1.1$  & $-2.1 \pm 2.0$ & $0.3 \pm 0.3$ \\
$-18$ & $1.2 \pm 1.3$ & $-2.0 \pm 2.1$ & $0.3 \pm 0.3$ \\
$-16$ & $1.7 \pm 1.4$ & $-1.8 \pm 2.1$ & $0.3 \pm 0.3$ \\
$-14$ & $1.6 \pm 1.4$ & $-1.3 \pm 2.2$ & $0.3 \pm 0.3$ \\
$-12$ & $1.7 \pm 1.5$  & $-0.9 \pm 2.2$ & $0.2 \pm 0.3$ \\
 $-10$ & $1.7 \pm 1.4$  & $-0.5 \pm 2.2$ & $0.1 \pm 0.3$ \\
 $-8$ & $1.5 \pm 1.4$  & $-0.1 \pm 2.3$ & $0.1 \pm 0.3$ \\
 $-6$ & $0.6 \pm 1.4$  & $0.3 \pm 2.4$ & $0.1 \pm 0.3$ \\
 $-4$ & $-0.4 \pm 1.2$ & $0.9 \pm 2.4$ & $0.0 \pm 0.2$ \\
 $-2$ & $-0.7 \pm 1.1$ & $1.3 \pm 2.5$ & $0.0 \pm 0.2$ \\
  $0$ & $-1.1 \pm 1.1$ & $1.8 \pm 2.5$ & $-0.1 \pm 0.2$ \\
  \hline \hline
\end{tabular}
\end{center}
\end{table}

\begin{table}[!t]
  \caption{Best-fit values of the branching fractions for unbound-state hypotheses.
  Uncertainties are statistical only.}
  \label{tab:best_fit_br_unbound}
  \begin{center}
\begin{tabular}{rrr}
  \hline \hline
 Mass difference (MeV) & $\Xi^0p$ ($\times 10^{-6}$) & $\Omega^-p$ ($\times 10^{-6}$) \\
 \hline
$2$ & $0.2 \pm 0.2$ & $-0.2 \pm 0.1$ \\
$4$ & $0.3 \pm 0.3$ & $-0.1 \pm 0.1$ \\
$6$ & $0.5 \pm 0.4$ & $0.0 \pm 0.2$ \\
$8$ & $0.6 \pm 0.4$ & $0.0 \pm 0.2$ \\
$10$ & $0.6 \pm 0.5$ & $0.0 \pm 0.2$ \\
$12$ & $0.6 \pm 0.5$ & $-0.1 \pm 0.2$ \\
$14$ & $0.6 \pm 0.5$ & $-0.2 \pm 0.2$ \\
$16$ & $0.6 \pm 0.5$ & $-0.2 \pm 0.2$ \\
$18$ & $0.7 \pm 0.5$ & $-0.2 \pm 0.2$ \\
$20$ & $0.7 \pm 0.6$ & $0.0 \pm 0.2$ \\
$22$ & $0.7 \pm 0.5$ & $0.0 \pm 0.2$ \\
$24$ & $0.6 \pm 0.5$ & $-0.1 \pm 0.3$ \\
$26$ & $0.5 \pm 0.4$ & $-0.2 \pm 0.3$ \\
$28$ & $0.5 \pm 0.4$ & $-0.3 \pm 0.3$ \\
$30$ & $0.5 \pm 0.4$ &  $-0.2 \pm 0.2$ \\
  \hline \hline
\end{tabular}
\end{center}
\end{table}





\end{document}